\documentclass[manuscript]{aastex}
\usepackage{url}
\shorttitle{SMA CO(3--2) \& OVRO CO(1--0) Observations of NGC
1068}
\shortauthors{Tsai et al.}

\begin{document}

\title{Interferometric CO(3--2) Observations toward the
    Central Region of NGC~1068}
\author{Mengchun Tsai\altaffilmark{1},
    Chorng-Yuan Hwang\altaffilmark{1},
    Satoki Matsushita\altaffilmark{2,3},
    Andrew J. Baker\altaffilmark{4},
    Daniel Espada\altaffilmark{5,6,2}
    }

\altaffiltext{1}{Institute of Astronomy, National Central University,
    Taiwan, R.O.C.}
\altaffiltext{2}{Academia Sinica Institute of Astronomy and
    Astrophysics, P.O.\ Box 23-141, Taipei 10617, Taiwan, R.O.C.}
\altaffiltext{3}{Academia Sinica Institute of Astronomy and
    Astrophysics, Joint ALMA Office, Av. El Golf 40, Piso 18, Las Condes, Santiago, Chile}
\altaffiltext{4}{Department of Physics and Astronomy, Rutgers,
    the State University of New Jersey, 136 Frelinghuysen Road,
    Piscataway, NJ 08854}
\altaffiltext{5}{Instituto de Astrof{\'i}sica de Andaluc{\'i}a
    - CSIC, Apdo. 3004, 18080 Granada, Spain}
\altaffiltext{6}{Harvard-Smithsonian Center for Astrophysics,
    60 Garden St., Cambridge, MA 02138}

\begin{abstract}
We present CO(3--2) interferometric observations of the central
region of the Seyfert 2 galaxy NGC~1068 using the Submillimeter
Array, together with CO(1--0) data taken with the Owens Valley Radio
Observatory Millimeter Array. Both the CO(3--2) and CO(1--0)
emission lines are mainly distributed within $\sim5$ arcsec of the
nucleus and along the spiral arms, but the intensity distributions
show differences; the CO(3--2) map peaks in the nucleus, while the
CO(1--0) emission is mainly located along the spiral arms. The
CO(3--2)/CO(1--0) ratio is about 3.1 in the nucleus, which is four
times as large as the average line ratio in the spiral arms,
suggesting that the molecular gas there must be affected by the
radiation arising from the AGN. On the other hand, the line ratios
in the spiral arms vary over a wide range from 0.24 to 2.34 with a
average value around 0.75, which is similar to the line ratios of
star-formation regions, indicating that the molecular gas is
affected by star formation. Besides, we see a tight correlation
between CO(3--2)/(1--0) ratios in the spiral arms and star formation
rate surface densities derived from \emph{Spitzer} $8~\micron$ dust
flux densities. We also compare the CO(3--2)/(1--0) ratio and the
star formation rate at different positions within the spiral arms;
both are found to decrease as the radius from the nucleus increases.
\end{abstract}


\section{INTRODUCTION}
\label{sect-intro}

The properties of circumnuclear molecular gas (CMG) near an active
galactic nucleus (AGN) can be influenced by the AGN itself. Some CMG
observations of Seyfert galaxies with a spatial resolution around
100 to 300 pc have shown that the intensities of CO(2--1) line
emission are about two times higher in temperature units than those
of CO(1--0) emission \citep[e.g.,][]{mat04,hsi08}. This is different
from the properties of molecular clouds in normal or star-forming
galaxies, which usually show brightness temperatures in higher-J
transitions similar to or lower than those in lower-J lines
\citep[e.g.,][]{dev94,mau99,dum01,oka07,mat09}. From the LVG model,
the high values of CO(2--1)/CO(1--0) arise when the emitting
molecular gas is in a low opacity environment and the resulting
molecular densities are larger than the critical densities of
CO(1--0) and CO(2--1) transitions. A possible origin for such
unusual behavior is that CMG near AGNs traces X-ray dominated
regions due to the strong radiation
\citep[e.g.,][]{mal96,koh01,use04,mei05,mei07}. Unlike far-UV
photons, X-ray photons have greater penetration lengths and are more
efficient in gas heating \citep{use04}. On the other hand, CMG can
also be affected by mechanical processes, such as gas entrainment by
jets \citep{mat07}.

Star formation in galaxies is well correlated with gas surface
density as embodied in the famous Schmidt-Kennicutt law
\citep{sch59,ken98}. In the circumnuclear regions (CNRs) and inner
structures of galaxies, the molecular gas usually has relatively
higher densities and temperatures than the molecular gas in galaxy
disks; in such environments, star formation is expected to be much
more vigorous \citep{ken98,tan00}. \citet{kt07} suggest that the
Schmidt-Kennicutt law might change slope when the averaged gas
density is close to the line critical density. We note that the
critical densities of CO(1--0) ($\sim10^3$~cm$^{-3}$) and CO(3--2)
$\sim 3\times10^4$~cm$^{-3}$ are comparable with the densities of
the CNRs of some nearby active galaxies, which are are around $300$
to $10^7$~cm$^{-3}$ \citep[e.g.,][]{mat04,sak07,per09}. It is
unclear whether or how the Schmidt-Kennicutt law would vary in the
CNRs of these galaxies when probed with CO(1--0) and CO(3--2)
emission.

Among nearby galaxies, NGC~1068 is the best-studied prototypical
Seyfert. It has a distance of 14.4~Mpc \citep[][;
$1\arcsec\sim70$~pc]{tul88} and is classified as (R)SA(rs)b in the
RC3 catalog \citep{vau91}, with an inner bar in the central kpc and
two tightly wound spiral arms starting from the tip of the bar
\citep{sco88,thr89}. The inner 2 kpc is rich in star formation
\citep{tel84}, and the active star forming regions are concentrated
along the spiral arms \citep[e.g.,][]{tel88,dav98,ems06}. Molecular
gas is also abundant along the spiral arms
\citep[e.g.,][]{mye87,pla91,kan92,hel95}, and a weak offset ridge of
emission along the leading side of the bar is also seen
\citep{hel95,sch00}, which is a typical molecular gas distribution
in barred galaxies.

The nucleus of NGC~1068 shows radio jets
\citep[e.g.,][]{wil83,gal96,gal04} and an ionization cone
\citep[e.g.,][]{pog88,mac94}. The nuclear optical spectrum has type
2 characteristics, but the polarized spectrum shows type 1 features,
reflecting the existence of an optically thick torus or disk around
the central massive black hole \citep{ant85}. Indeed, a sub-parsec
scale ionized gas disk perpendicular to the radio jets has been
observed \citep{gal97,gal04}, with a molecular gas (maser) torus
\citep{gre96} or disk \citep{gal01} located outside of the ionized
gas disk. This maser torus/disk may be surrounded by a pc-scale warm
dust torus \citep{jaf04}. Outside this structure, the existence of a
warped molecular gas disk is suggested by interferometric CO(2--1)
observations \citep{bake98,sch00}, but warm molecular gas kinematics
dominated by irregular (infalling) motions have been observed in
$2.12~\micron$ 1--0 S(1) molecular hydrogen emission \citep{mul09}.
Interferometric observations of molecular gas emission with angular
resolutions at $0.\arcsec$5--$2\arcsec$ also indicated non-circular
motions in the central $\sim$100 pc \citep{krip11}. The intensity
ratios of various molecular species exhibit peculiar values,
including very high HCN/CO and HCN/HCO$^{+}$ ratios
\citep{jac93,tac94,koh01,use04,krip08,per09}, suggesting that the
molecular gas very close to the Seyfert 2 nucleus is irradiated by
strong X-ray emission \citep{use04,koh05,koh08,gar10}.

The central region of NGC~1068 has been mapped in different CO
tranistion lines by both single-dish and interferometric
observations. Among the former, it has been observed by
\citet[CO(1--0), FCRAO)]{sco83}, \citet[CO(1--0), Nobeyama]{kan89},
\citet[CO(1--0) and CO(2--1), IRAM]{pla89}, \citet[CO(1--0),
FCRAO]{you95}, \citet[CO(2--1) and CO(3--2), JCMT]{pap99}, and
\citet[CO(1--0), CO(2--1), CO(3--2), and CO(4--3), IRAM and
JCMT]{israel09}. With interferometers, it has been observed by
\citet[CO(1--0), OVRO]{pla91}, \citet[CO(1--0), NMA]{kan92},
\citet[CO(1--0), BIMA]{hel95}, \citet[CO(1--0) and CO(2--1),
IRAM]{sch00}, \citet[CO(1--0), BIMA]{hel03}, and \citet[CO(2-1) and
CO(3-2), SMA/PdBI]{krip11}. In this paper, we present
interferometric observations of the CO(3--2) line in the central 2
kpc of NGC~1068 using the Submillimeter Array (SMA). We also show
interferometric observations of the CO(1--0) transition, with $uv$
sampling similar to that of the CO(3--2) data, from the Owens Valley
Radio Observatory (OVRO) Millimeter Array. We describe our
observations in Sect.~\ref{sect-obs} and show the overall molecular
gas distributions in Sect.~\ref{sect-res-dist}. Line ratios,
molecular gas masses, and kinematics are presented and discussed in
Sect.~\ref{sect-res-ratio}, \ref{sect-res-mass}, and
\ref{sect-res-rot}, respectively. The relation between line ratios
and star formation is discussed in Sect.~\ref{sect-res-sfr}, and a
summary is in Sect.~\ref{sect-sum}.

\section{OBSERVATIONS AND DATA REDUCTION}
\label{sect-obs}

\subsection{SMA CO(3--2) Observations}
\label{sect-obs-sma}

We used the SMA to acquire three CO(3--2) datasets on August 13,
15, and 23, 2005. The zenith opacity at 225 GHz was about 0.07 in
August 13 and 15 and 0.06 in August 23. Six out of eight 6-m
antennas were used in the compact configuration. The receivers
were tuned to the redshifted CO(3--2) line ($344.493$~GHz).
Correlators were set to cover a velocity range of
$\sim1700$~km~s$^{-1}$ ($\sim2$~GHz bandwidth) and configured to
have a velocity resolution of $\sim0.7$~km~s$^{-1}$ ($=0.8125$~MHz
frequency resolution). The SMA antenna primary beam has a half
power beam width (HPBW) of $\sim36\arcsec$ ($\sim 2.5$~kpc in
NGC~1068) at $345$~GHz. A 7-pointing mosaic (pointings separated
by $18\arcsec$ and hexagonal in shape) was observed in order to
image the inner $72\arcsec$ (5 kpc) of NGC~1068. The phase center
was set at R.A.$ =2h42m40.798s $ and decl.$=-0d00m47.938s$
(J2000), corresponding to the AGN position \citep{mux96}.

We performed the data reduction following the standard processes
outlined in the SMA
cookbook\footnote{\url{http://cfa-www.harvard.edu/~cqi/mircook.html}}.
The visibilities were first calibrated with the IDL-based MIR
package \citep{sco93} as modified for the SMA. We used $0423-013$
as the amplitude and phase calibrator to track phase and gain
variations and used 3C~454.3 for bandpass and flux calibrations.

We calibrated our observations by linearly interpolating the flux
densities of the quasar 3C~454.3, which were $26.48\pm1.33$~Jy on
August 12 and $18.90\pm0.95$~Jy on August 25, according to the SMA
calibrator
list\footnote{\url{http://sma1.sma.hawaii.edu/callist/callist.html}}.
The expected flux densities of~3C 454.3 were about $25\pm 1.3$
(August 13-15), and $20\pm 1.0$ (August 23) during the observations.
Mars and Uranus were also observed during the observations. However,
Mars was resolved by the SMA observations, and Uranus was only
observed on August 13 and 15. After performing the flux calibration
with 3C~454.3, we compared the fluxes of Uranus and Mars to check
the accuracy of our flux calibration. The uncertainty in our flux
scale is estimated to be $\sim 10\%$.

We combined the visibilities from all the mosaic observations and
applied the mapping task INVERT in MIRIAD to produce ``dirty''
images with a velocity resolution of 10~km~s$^{-1}$. Primary beam
correction was taken into account in the mosaic mode of the INVERT
process. We used natural weighting in the mapping process in order
to have the best sensitivity. We deconvolved the dirty images with
the MOSSDI and MOSMEM packages in MIRIAD, and produced moment maps
and spectra for further analysis. The resulting spatial resolution
was $2\farcs29\times2\farcs00$ ($\sim160\times140~$pc$^2$) with a
position angle of $-88\fdg8$.

\subsection{OVRO CO(1--0) Observations}
\label{sect-obs-ovro}

We observed the CO(1--0) transition in NGC\,1068 (adopted pointing
center: $\alpha_{\rm J2000} = $ 02:42:40.7 and $\delta_{\rm J2000} =
$ $-$00:00:47.7) between April and September 1995 using the OVRO
millimeter array \citep{padi91,scot93}.  The array comprised six
10.4\,m antennas, which during our observations were deployed in
three configurations providing a total of 40 distinct baselines.  We
configured the array's digital correlator \citep{padi93} to provide
112 contiguous frequency channels, each Hanning-smoothed to $4\,{\rm
MHz}$ ($10.4\,{\rm km\,s^{-1}}$) resolution.  We calibrated the data
within the OVRO millimeter array database using the MMA package
\citep{sco93}.  Paired integrations on J0339$-$017, interleaved with
observations of NGC\,1068 every 30-40 minutes, were used to correct
for phase and gain variations; for passband calibration, we used
3C273, 3C454.3, and/or 0528+134. Our flux scale was determined by
comparing J0339$-$017 with Uranus or Neptune in light of standard
brightness temperature models for the latter \citep{muhl91,orto86},
or by bootstrapping from archived observations of the planets (and
bright, frequently-observed quasars) obtained with similar
elevations and coherences. From repeatibility of flux measurements
on 1--2 week timescales, we estimate that the uncertainty in our
flux scale is $\sim 10\%$.

After editing for quality in the Difmap package \citep{shep97}, we
were left with 11.4 hours of on-source data.  To eliminate
contamination by 3\,mm continuum emission associated with
NGC\,1068's jet, we subtracted a $uv$-plane model based on the
line-free channels at the ends of the recorded bandwidth. The
line-free $uv$ data were then mapped using the IMAGR task in the
NRAO AIPS package, with moderately robust weighting giving a
synthesized beam of $3\farcs46 \times 2\farcs56$ (242\,pc $\times$
179\,pc) at PA $67\fdg5$. For deconvolution, we adopted a single
clean box for all channels that enclosed all of the emission in the
zeroth moment map. A slightly different reduction of these data was
first discussed by \citet{bake98}.  Before analyzing moment maps and
spectra, we applied a primary beam correction appropriate for the
$\sim 56^{\prime\prime}$ OVRO HPBW.


\section{CO Distribution}
\label{sect-res-dist}

\begin{figure}
\includegraphics[angle=-90,width=.5\textwidth]{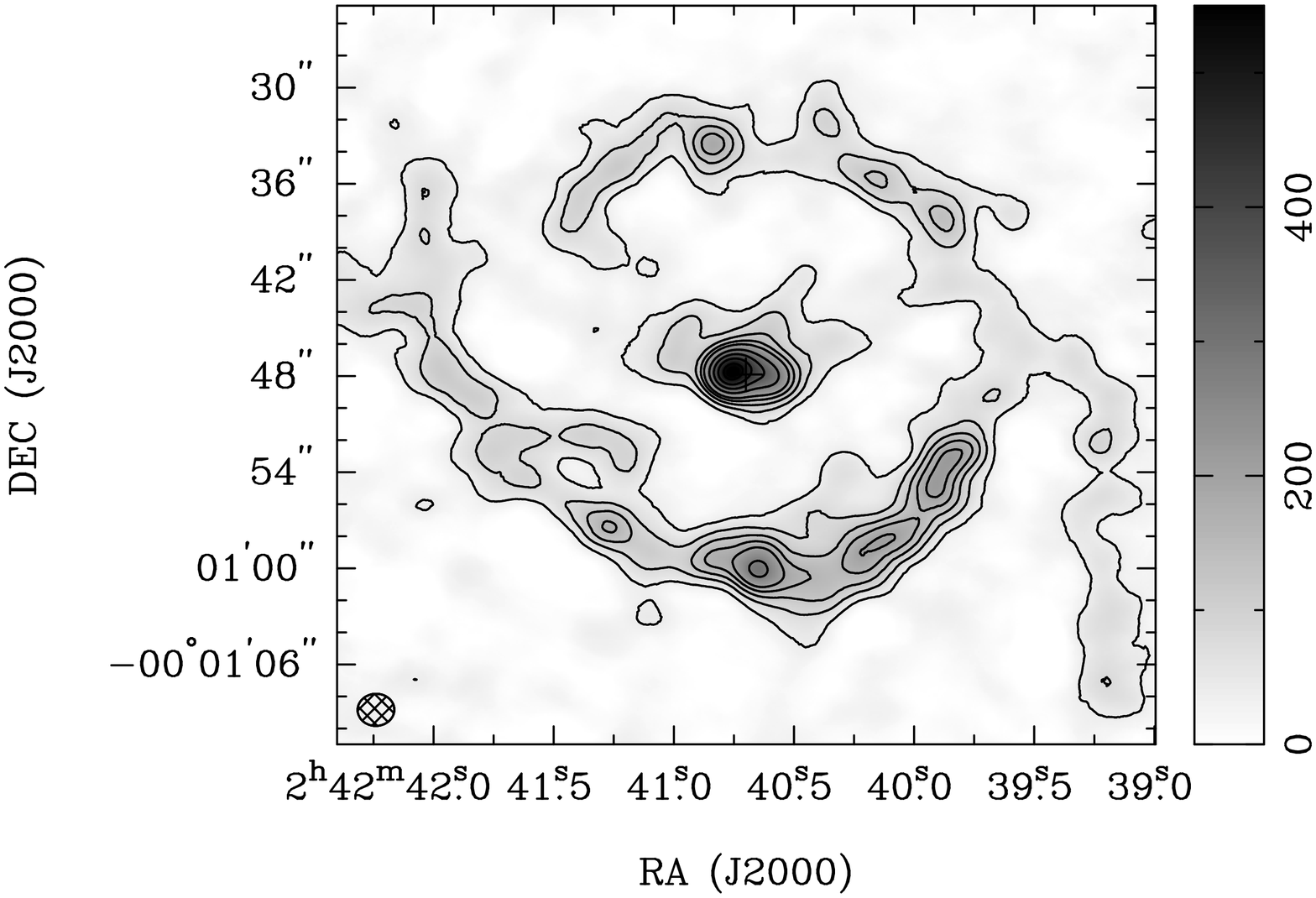}
\includegraphics[angle=-90,width=.5\textwidth]{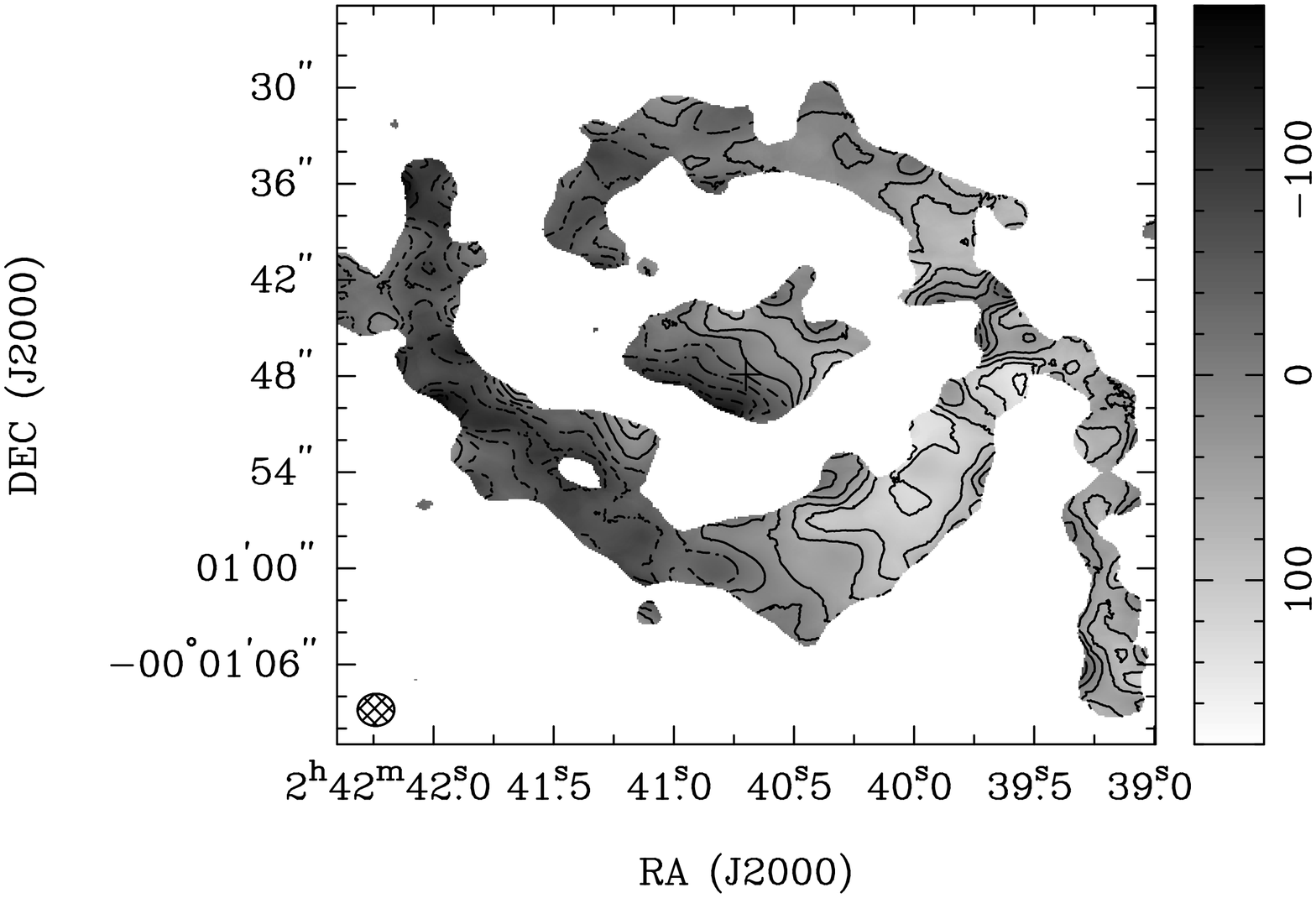}
\caption{$Left$: CO(3--2) integrated intensity (moment 0) map of
the central region of NGC~1068. The grayscale range is shown in
the wedge at right in units of Jy~beam$^{-1}$~km~s$^{-1}$. The
contour levels are 10, 20, 30, 40, 50, 70, 90, 110, and 130
$\times\sigma$, where $1\sigma= 4.0$~Jy~beam$^{-1}$~km~s$^{-1}$.
The synthesized beam is $2\farcs29\times2\farcs00$
($\sim160\times140\rm~pc^2$) with a P.A.\ of $-88\fdg0$, which is
shown in the bottom left corner. The cross indicates the galactic
center determined from the 5~GHz and 22~GHz radio continuum data
\citep{mux96}. $Right$: CO(3--2) intensity-weighted velocity field
(moment 1) map. The grayscale range is shown in the wedge at right
from -180 to 180 km~s$^{-1}$. The contour levels range from $-150$
to $+150$~km~s$^{-1}$ with a 30~km~s$^{-1}$ interval.
\label{fig1}}
\end{figure}

In Figure~\ref{fig1}, we show the integrated intensity and velocity
maps of the CO(3--2) line for the central region of NGC~1068.
CO(3--2) emission appears in the nucleus and along the two spiral
arms. The CO(3--2) distribution in the nucleus is elongated in an
east-west direction, and the strongest peak in our data is located
$\sim1\arcsec$ east of the nucleus. The distribution of the CO(3--2)
emission is roughly consistent with that of previous CO(1--0)
observations \citep[e.g.,][]{hel95,pla91,kan92,sch00} except that
the nucleus is relatively brighter in the CO(3--2) image.

The CO(3--2) intensity distributions in the two spiral arms are
different. The southern arm is brighter than the northern arm. The
second strongest peak in our map is located $\sim12\arcsec$ south of
the nucleus, which is on the southern arm. The locations of these
two strongest peaks in the map can explain the position of the peak
intensity in an early CO(3--2) single-dish image \citep{pap99},
which showed a peak intensity offset south of the nucleus. The
missing flux in the central $14\arcsec$ region is about 20\% when
comparing our image with a JCMT observation of the central region of
NGC~1068 \citep{israel09}. On the other hand, if we compare our data
with an area-averaged CO(3-2) spectrum over a circular area of
$30\arcsec$ radius \citep{pap99}, we find that our map only recovers
$\sim50\%$ of the total flux. The missing flux might seem
surprisingly large; however, \citet{pap99} have showed that the
total CO(3--2) flux is actually dominated by a warm diffuse gas
phase, which is highly excited and not virialized. Besides, similar
extended structures of CO(3--2) emission are also observed in the
central regions of other nearby AGNs, such as M51 \citep{mat04}.
Most of this diffuse gas is likely resolved out and not detectable
in our interferometer map.

\begin{figure}
\includegraphics[angle=-90,width=.5\textwidth]{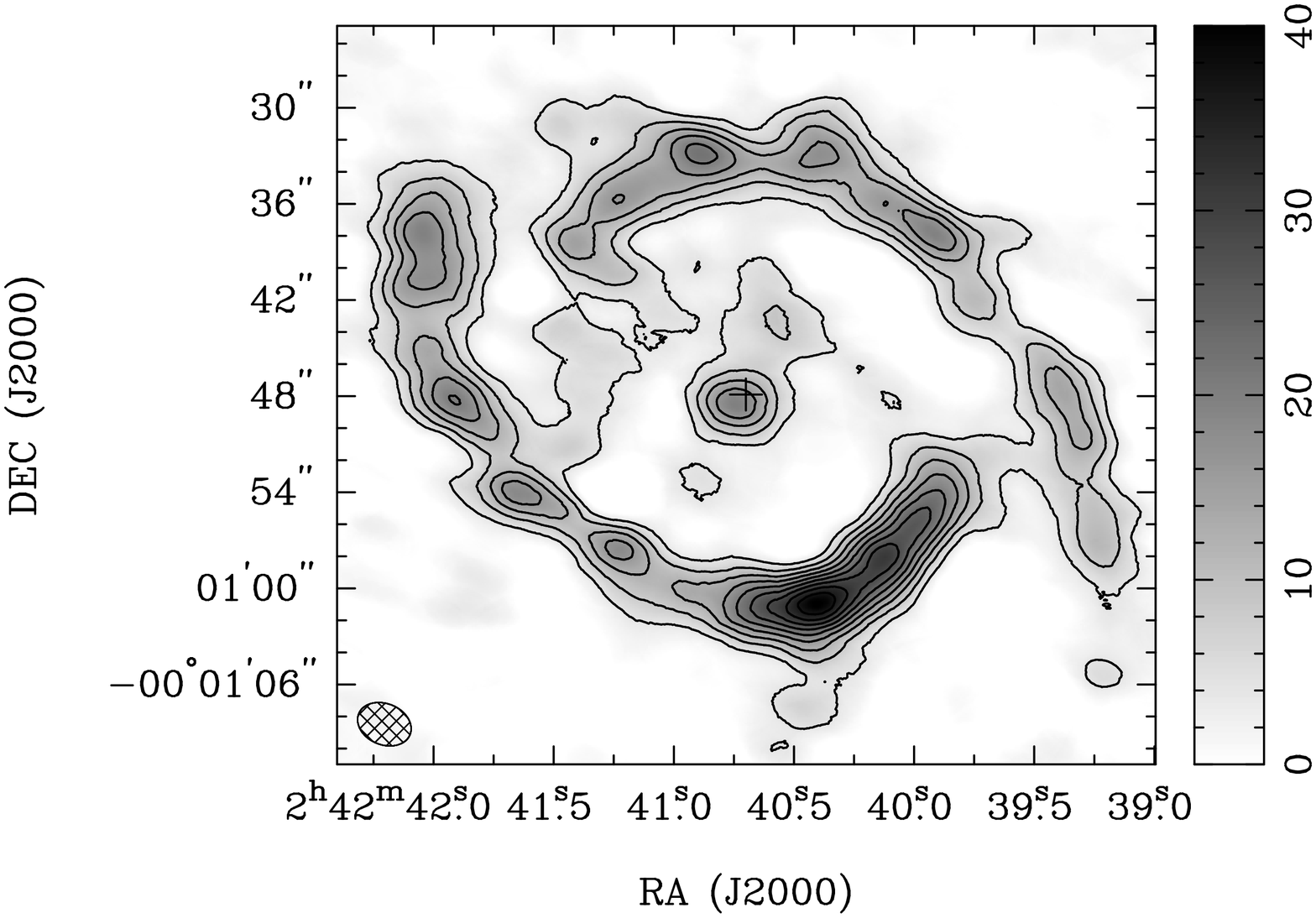}
\includegraphics[angle=-90,width=.5\textwidth]{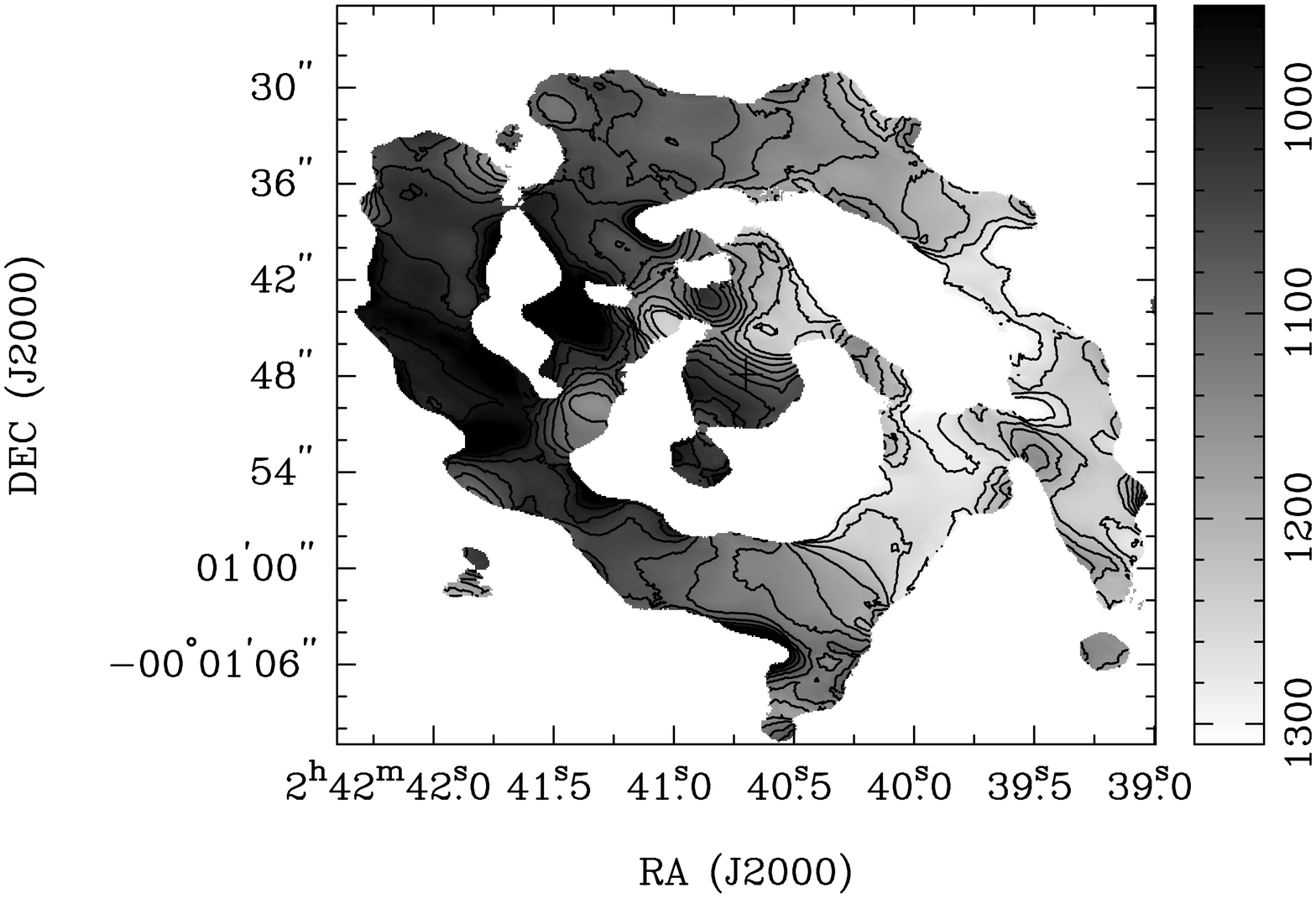}
\caption{$Left$: CO(1--0) integrated intensity (moment 0) map of
the central region of NGC~1068. The grayscale range is shown in
the wedge at right in units of Jy~beam$^{-1}$~km~s$^{-1}$. The
contour levels are multiplies of $5\sigma$, where $1\sigma=
0.81$~Jy~beam$^{-1}$~km~s$^{-1}$. The synthesized beam is
$3\farcs46\times2\farcs56$ with a P.A.\ of $67\fdg5$, which is
shown in the bottom left corner. The cross is the same as in
Fig.~\ref{fig2}. $Right$: CO(1--0) intensity weighted velocity
field (moment 1) map of the central region of NGC~1068. The
grayscale range is shown in the wedge at the right, from -180 to
180 km~s$^{-1}$. The contour levels range from $-150$ to
$+150$~km~s$^{-1}$ with a 30~km~s$^{-1}$ interval. \label{fig2}}
\end{figure}

The CO(1--0) integrated intensity and velocity maps are shown in
Figure~\ref{fig2}. As in the CO(3--2) map, CO(1--0) emission is
located in the nucleus and along the two spiral arms, but the
strongest peak is located at the southern arm, and the peak
intensity at the core region is relatively weak. Our CO(1--0) image
is  in general agreement with previous interferometric CO(1--0)
observations \citep{pla91, kan92, sch00, hel03}. However, the BIMA
image seems to show more emission at interarm regions. The
difference between our results and the BIMA image might be due to
the fact that the BIMA image have included modelled visibilities
derived from single-dish NRAO observations and might contain
large-scale emission that would be resolved out by our
interferometric observations.

The corresponding missing flux of CO(1--0) in the core region is
about 20\% comparing with the BIMA results of \citet{hel03}. For the
spiral regions, the situation is more complicated. We have estimated
the missing fluxes for several selected spiral regions (see Figure
3) and obtained diverse results. For example, the missing flux for
R4 is about 70\% and for R17 is about 38\%; however, we detect more
flux than BIMA for R15, so there should be no missing flux for R15.
We note that R4 is an inter-arm region, which is dominated by
diffuse emission, while R15 is around a compact structure peak and
R17 is at the boundary of an arm structure. These different results
might be caused by different $uv$ coverage of the BIMA and our
observations.

The overall CO(1--0) velocity field is similar to that of CO(3--2).
Both have a rotation axis with P.A. about $-30\degr$ and a rotation
velocity of 120~km~s$^{-1}$ in the nucleus. In the spiral arms, both
kinematic major axes run from east to west with a rotation velocity
of 180~km~s$^{-1}$.

\section{Discussion}
\label{sect-dis}

\subsection{CO Line Ratios}
\label{sect-res-ratio}

\begin{figure}
\includegraphics[angle=-90,width=.9\textwidth]{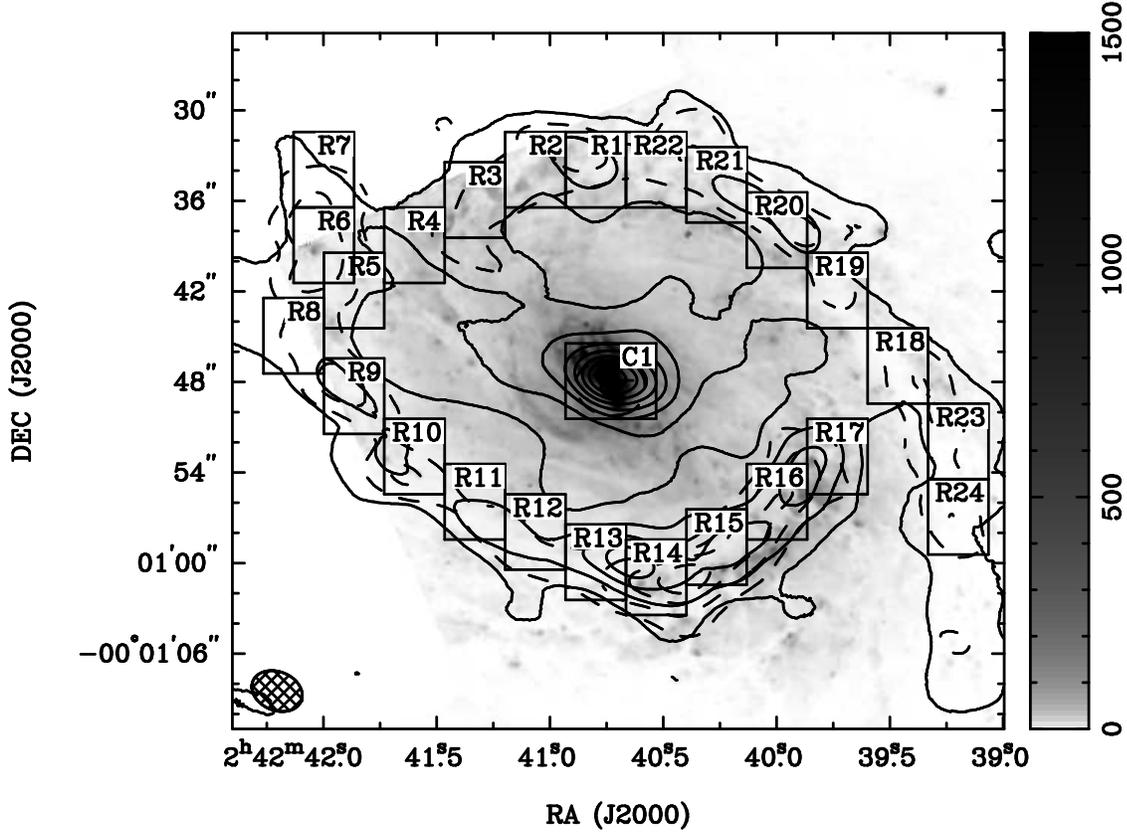}
\caption{ Integrated intensity maps of CO(3--2) (solid
    contours) and CO(1--0) (dashed contours), overlaid on the
    continuum-subtracted \emph{HST} F658N image (greyscale) of the central
    region of NGC~1068.
    Solid contour levels for CO(3--2) are 10, 30, 50, 70, 90, 110, and
    130 $\times$ 5.3~Jy~beam$^{-1}$~km~s$^{-1}$, and dashed contour
    levels for CO(1--0) are 2, 4, 6, 8 $\times$ 5~Jy~beam$^{-1}$~km~s$^{-1}$.
    The CO(3--2) and CO(1--0) data are matched to the same $uv$ range
    and have the same synthesized beam size of $3\farcs46\times2\farcs56$ ($\sim242\times179\rm~pc^2$) with
    a P.A.\ of $67\fdg5$, which is shown in the bottom-left corner of the image.
    Cross is the same as in Figure~\ref{fig1}.
    We also plot 25 boxes (C1 and R1 -- R24) that are used to calculate the line ratios. \label{fig3}}
\end{figure}

Before deriving line ratios at various regions, we first matched
the $uv$ range between our CO(3--2) and CO(1--0) datasets. In
Figure~\ref{fig3}, we overlay the CO(3--2) (solid contours) and
CO(1--0) (dashed contours) intensity distributions. In this image,
the shortest $uv$ length is set to 16~k$\lambda$, and the image
resolutions are convolved into the same resolution
($3\farcs46\times2\farcs56$ with a P.A.\ of $67\fdg5$). As
mentioned above, the overall distributions of both CO(3--2) and
CO(1--0) are very similar, but some of the intensity peaks have
shifted positions with respect to each other. The most obvious
example is in the southern spiral arm, about $12\arcsec-14\arcsec$
south of the nucleus, where the CO(3--2) peak is located in the
inner part of the spiral arm, but the CO(1--0) peak is shifted
toward the outer edge of the spiral arm.

We divide the central region of NGC~1068 into 25 regions covering
the nucleus and spiral arms, as shown in Figure~\ref{fig3} (C1 and
R1 -- R24). The size of each region is $4\arcsec\times5\arcsec$
except C1, which is $6\arcsec\times5\arcsec$. The intensity scale is
then converted to a brightness temperature scale, and the
CO(3--2)/CO(1--0) brightness temperature ratios, $R_{31}$, are
derived for each region using the MIRIAD task MATHS. Area-averaged spectral peak
and integrated brightness temperatures and $R_{31}$
for each region are shown in Table~\ref{tab-ratio}.

\begin{figure}
\includegraphics[angle=0,width=\textwidth]{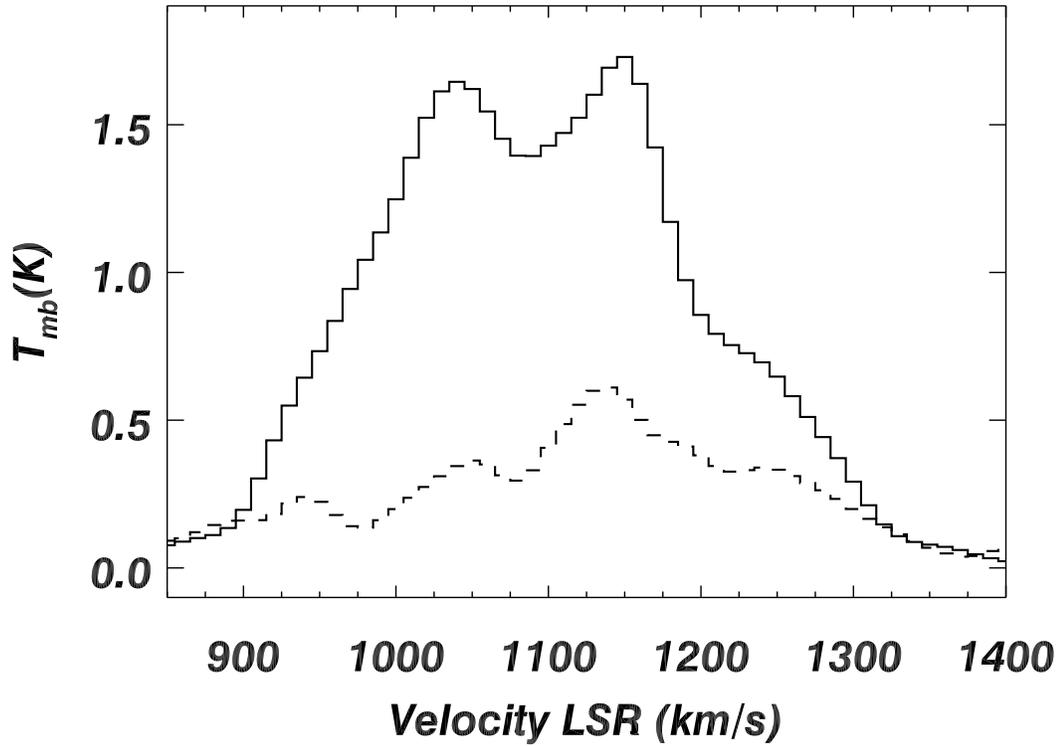}
\caption{Area-averaged spectra within the C1 region. The solid line shows the spectrum of
CO(3--2), and the dash line shows the spectrum of CO(1--0).
The uncertainties per channel are about 0.01~K for the CO(3--2) spectrum and 0.02~K for the CO(1--0) one. \label{fig4}}
\end{figure}

The central core region C1 exhibits a very high integrated intensity
ratio with $R_{31}=3.12\pm0.03$ and a spectra peak ratio of
$2.83\pm0.10$. Figure~\ref{fig4} shows the spectra of C1.
The CO(3--2)/CO(1--0) line ratio is slightly smaller than the results of \citet{krip11},
which are about 4--6.
However, the observations of \citet{krip11} have a much higher angular resolution and might
contain relatively more contribution from the circumnuclear region of the AGN, while our result
is an average value over a much larger area and are likely to include emission from
outside regions. In fact, if we only consider
the line ratio within the brightest beam at the center,
we would obtain a line ratio of
$\sim4.6$, which is similar to the results of \citet{krip11}.

On the other hand, the spiral arm regions R1 -- R24 have a wide
range of integrated intensity ratios of 0.24 -- 2.34 and an average
value of 0.75 with standard deviation 0.47. We note that the wide
range of the intensity ratios might be caused by the different
spatial distributions of the CO(1--0) and CO(3--2) emission due to
varying physical conditions of the molecular gas. There is no
obvious ratio difference between the northern and southern arms; the
average ratio of the northern arm regions (R1 -- R4, R18 -- R24) is
0.8 and that of the southern arm regions (R5 -- R17) is 0.7.
However, there is a difference between inner arm regions and outer
arm regions; the average ratio of the inner arm regions (R1 -- R4,
R9 -- R17, and R19 -- R22) is $0.88\pm0.14$, while the outer arm
regions (R5 -- R8, R18, R23, and R24) show $0.42\pm0.05$. This
difference indicates that there is a large-scale gradual decrease of
the line intensity ratio from the inner radii to the outer radii.
The radial change of molecular line ratios, and therefore of the
physical conditions of the molecular gas, has also been detected in
the Milky Way and other galaxies
\citep{tur93,aal94,sak94,sak97,pet00,pag01,mat10}, suggesting that
this trend is common.

The line ratios $R_{31}$ in the spiral arm regions are similar to
those observed in the centers of nearby normal and starburst
galaxies. \citet{dev94} observed seven nearby starburst galaxies and
found that the ratios are in the range of 0.5 -- 1.4 with an average
ratio of $0.64\pm0.06$. A survey toward the centers of 28 nearby
star-forming galaxies showed that most of their $R_{31}$ are within
the range of 0.2 -- 0.7 \citep{mau99}. On the other hand, the value
of $R_{31}$ in the nucleus is much larger than those observed in the
nuclei of the nearby galaxies mentioned above, albeit similar to
that observed in the nucleus of the Seyfert 2 galaxy M51. The inner
$uv$-truncated $R_{31}$ of the central molecular core of M51 also
has a very high value of $\sim5.2\pm1.7$ within a beam size of
$3\farcs9\times2\farcs6$ or 160~pc $\times$ 110~pc \citep[missing
flux corrected $R_{31}$ is $\geq 1.9\pm0.2$;][]{mat04}. If we only
correct the 20\% missing flux of the CO(1--0) emission in the core
region of NGC~1068 and ignore the missing flux of the CO(3--2)
following the method of \citet{mat04}, the $R_{31}$ ratio of the
core region will become 2.50 instead of 3.12. This suggests that the
physical conditions of molecular gas around the Seyfert 2 AGN in
NGC~1068 are very different from those in the centers of nearby
star-forming and starburst galaxies, but might be similar to those
around the Seyfert 2 nucleus of M51.

\begin{deluxetable}{cccccccccc}
\rotate \tabletypesize{\scriptsize} \tablecaption{Peak and
integrated brightness temperature of CO(3--2) and CO(1--0), and
the CO(3--2)/CO(1--0) line ratios, in different areas.
\label{tab-ratio}} \tablehead{
    \colhead{Area} & \colhead{Peak CO(3--2)} & \colhead{Peak CO(1--0)}
        & \colhead{Peak $R_{31}$}
        & \colhead{$\int T_{\rm B}(\rm{CO}$~$J = 3 - 2)~dv$}
        & \colhead{$\int T_{\rm B}(\rm{CO}$~$J = 1 - 0)~dv$}
        & \colhead{$R_{31}$} & \colhead{X offset} & \colhead{Y offset} \\
        & \colhead{[K]} & \colhead{[K]} &
        & \colhead{[K~km~s$^{-1}$]} & \colhead{[K~km~s$^{-1}$]} &
    & \colhead{[arcsec]} & \colhead{[arcsec]}
    }
\startdata
  C1 & $1.73\pm0.01$ & $0.61\pm0.03$ & $2.83\pm0.10$
     & $ 438.5\pm 0.7$ & $ 140.6\pm  1.5$ & $ 3.12\pm0.03$ & -1 & 0\\
  R1 & $1.80\pm0.03$ & $1.55\pm0.06$ & $1.16\pm0.05$
     & $ 103.76\pm 0.97$ & $  44.3\pm  1.8$ & $ 2.34\pm0.10$ & 0 & 14\\
  R2 & $1.53\pm0.05$ & $1.90\pm0.08$ & $0.80\pm0.04$
     & $ 104.97\pm 1.54$ & $ 133.4\pm  2.8$ & $ 0.79\pm0.02$ & 4 & 14\\
  R3 & $1.29\pm0.06$ & $1.98\pm0.05$ & $0.65\pm0.03$
     & $ 100.06\pm 2.07$ & $ 149.8\pm  1.8$ & $ 0.67\pm0.02$ & 8 & 12\\
  R4 & $0.43\pm0.03$ & $0.73\pm0.04$ & $0.58\pm0.05$
     & $   6.26\pm 0.81$ & $  26.6\pm  1.2$ & $ 0.24\pm0.03$ & 12 & 9\\
  R5 & $0.78\pm0.04$ & $1.92\pm0.11$ & $0.41\pm0.03$
     & $  56.74\pm 1.39$ & $  90.0\pm  3.8$ & $ 0.63\pm0.03$ & 16 & 6\\
  R6 & $1.54\pm0.05$ & $3.65\pm0.15$ & $0.42\pm0.02$
     & $  78.95\pm 1.45$ & $ 131.1\pm  4.4$ & $ 0.60\pm0.02$ & 18 & 9\\
  R7 & $0.71\pm0.04$ & $1.97\pm0.14$ & $0.36\pm0.03$
     & $  25.72\pm 1.23$ & $  93.0\pm  4.4$ & $ 0.28\pm0.02$ & 18 & 14\\
  R8 & $0.84\pm0.03$ & $1.80\pm0.08$ & $0.47\pm0.03$
     & $  28.85\pm 1.09$ & $  94.4\pm  2.5$ & $ 0.31\pm0.01$ & 20 & 3\\
  R9 & $1.35\pm0.03$ & $3.63\pm0.08$ & $0.37\pm0.01$
     & $ 108.83\pm 1.11$ & $ 199.9\pm  2.9$ & $ 0.54\pm0.01$ & 16 & -1\\
 R10 & $1.49\pm0.07$ & $2.20\pm0.08$ & $0.68\pm0.04$
     & $  62.03\pm 1.92$ & $ 111.0\pm  2.2$ & $ 0.56\pm0.02$ & 12 & -5\\
 R11 & $1.49\pm0.07$ & $1.94\pm0.03$ & $0.77\pm0.04$
     & $  94.26\pm 2.63$ & $ 172.2\pm  1.1$ & $ 0.55\pm0.02$ & 8 & -8\\
 R12 & $1.74\pm0.06$ & $3.02\pm0.04$ & $0.57\pm0.02$
     & $ 112.76\pm 2.01$ & $ 176.3\pm  1.4$ & $ 0.64\pm0.01$ & 4 & -10\\
 R13 & $1.93\pm0.09$ & $1.66\pm0.07$ & $1.16\pm0.07$
     & $ 194.32\pm 3.64$ & $ 139.4\pm  2.8$ & $ 1.39\pm0.04$ & 0 & -12\\
 R14 & $2.46\pm0.09$ & $3.40\pm0.03$ & $0.72\pm0.03$
     & $ 252.08\pm 3.58$ & $ 192.5\pm  1.3$ & $ 1.31\pm0.02$ & -4 & -13\\
 R15 & $2.68\pm0.08$ & $3.18\pm0.09$ & $0.84\pm0.04$
     & $ 239.52\pm 3.24$ & $ 374.4\pm  3.4$ & $ 0.64\pm0.01$ & -8 & -11\\
 R16 & $3.46\pm0.11$ & $4.80\pm0.18$ & $0.72\pm0.04$
     & $ 211.39\pm 3.69$ & $ 301.1\pm  6.0$ & $ 0.70\pm0.02$ & -12 & -8\\
 R17 & $1.67\pm0.01$ & $1.68\pm0.06$ & $0.99\pm0.04$
     & $ 106.98\pm 0.22$ & $ 141.2\pm  2.3$ & $ 0.76\pm0.01$ & -16 & -5\\
 R18 & $0.79\pm0.01$ & $2.47\pm0.07$ & $0.32\pm0.01$
     & $  56.68\pm 0.62$ & $ 159.4\pm  2.8$ & $ 0.36\pm0.01$ & -20 & 1\\
 R19 & $1.28\pm0.05$ & $1.72\pm0.08$ & $0.75\pm0.04$
     & $  49.41\pm 1.41$ & $  75.5\pm  2.4$ & $ 0.65\pm0.03$ & -16 & 6\\
 R20 & $1.94\pm0.07$ & $3.19\pm0.06$ & $0.61\pm0.02$
     & $ 115.44\pm 2.06$ & $ 168.9\pm  1.8$ & $ 0.68\pm0.01$ & -12 & 10\\
 R21 & $1.83\pm0.05$ & $1.66\pm0.06$ & $1.10\pm0.05$
     & $  97.04\pm 1.48$ & $  81.3\pm  1.8$ & $ 1.19\pm0.03$ & -8 & 13\\
 R22 & $1.10\pm0.04$ & $0.47\pm0.04$ & $2.32\pm0.22$
     & $  62.59\pm 1.26$ & $  46.3\pm  1.3$ & $ 1.35\pm0.05$ & -4 & 14\\
 R23 & $1.02\pm0.05$ & $2.72\pm0.07$ & $0.38\pm0.02$
     & $  45.12\pm 1.70$ & $  94.9\pm  2.1$ & $ 0.48\pm0.02$ & -24 & -4\\
 R24 & $0.96\pm0.03$ & $2.95\pm0.16$ & $0.33\pm0.02$
     & $  39.06\pm 0.86$ & $ 126.5\pm  4.5$ & $ 0.31\pm0.01$ & -24 & -9\\
\enddata
\tablecomments{Column (1): Region name. Column (2): Peak brightness
temperature of CO(3--2). Column (3): Peak brightness temperature of
CO(1--0). Column (4): Line ratio of peak brightness temperature of
CO(3--2)/CO(1--0). Column (5): Integrated brightness temperature of
CO(3--2). Column (6): Integrated brightness temperature of CO(1--0).
Column (7): Line ratio of integrated brightness temperature of
CO(3--2)/CO(1--0). The errors Column (8,9): Position offset of
region center from the phase center.  Errors were estimated from the
statistical errors of the spectra; uncertainties of our flux
calibration (about 10\% for both CO(1--0) and CO(3--2) lines) and
missing fluxes are not included.}

\end{deluxetable}

\subsection{Molecular Gas Column Density and Mass}
\label{sect-res-mass}

Because CO(1--0) traces the bulk of the molecular gas, we can
calculate the molecular gas column density and mass from the
CO(1--0) integrated intensities. The column density of the
molecular hydrogen can be estimated with a conversion factor,
$X_{\rm CO}$,
\begin{equation}
N_{\rm H_2} = X_{\rm CO} \int T_{\rm B}({\rm CO}~J = 1 - 0)~dv
    ~{\rm [cm^{-2}]},
\end{equation}
where $X_{\rm CO}\cong0.4\times10^{20}$~cm$^{-2}$~(K km
s$^{-1}$)$^{-1}$ for circumnuclear molecular gas
\citep{wil95,mau96,weis01,esp09} and $X_{\rm
CO}\cong3\times10^{20}$~cm$^{-2}$~(K~km~s$^{-1}$)$^{-1}$ for spiral
arm regions \citep{sol87,you91}. The column density $N_{\rm H_2}$
for each region is calculated using $\int T_{\rm B}({\rm CO}~J = 1 -
0)~dv$ shown in Table~\ref{tab-ratio}, and the derived values are
shown in Table~\ref{tab-sfr}. We also derive the molecular gas mass,
$M_{\rm H_2}$, which is displayed in Table~\ref{tab-sfr}. The total
missing flux of CO(1--0) within the central circular area of
$30\arcsec$ radius is about 35\% \citep{pap99}. However, the missing
flux is dominated by extended structures, so the true missing fluxes
in the core and the compact spiral structures should be smaller. For
example, the missing flux of CO(1--0) in the core region is about
20\% as discussed in Sect.~\ref{sect-res-dist}. We thus expect that
the errors of the column densities caused by the missing flux are
less than 35\%.

\begin{deluxetable}{ccccccccc}
\tabletypesize{\scriptsize}
\tablecaption{Molecular gas column density and mass, $8~\mu$m dust
    and H$\alpha$ + [N II] emission, and star
    formation rate surface density.
    \label{tab-sfr}}
\tablehead{
    \colhead{Area} & \colhead{$N_{\rm H_2}$}
        & \colhead{$M_{\rm H_2}$}
        & \colhead{$f_{\rm 8\micron, dust}$}
        & \colhead{$f_{\rm H\alpha+[N II]}$}
        & \colhead{$\Sigma_{\rm SFR}$} \\
        & \colhead{$\times10^{21}$ [cm$^{-2}$]}
        & \colhead{$\times10^{6}$ [$M_\odot$]}
        & \colhead{[MJy~sr$^{-1}$]}
        & \colhead{$\times10^{4}$ [eps~pixel$^{-1}$]\tablenotemark{a}}
        & \colhead{[$M_\odot$~yr$^{-1}$~kpc$^{-2}$]}
    }
\startdata
  C1 & $  5.6\pm0.3$ & $ 12.3\pm0.8$ & $1.05\times10^{4}\pm6.4$ & 73.5 & $62.4\pm0.04$\\
  R1 & $  13.3\pm0.5$ & $ 19.5\pm0.8$ & $238\pm6.4$ & 1.51 & $1.41\pm0.04$\\
  R2 & $  40.0\pm0.8$ & $ 58.6\pm1.2$ & $260\pm6.4$ & $>1.87$\tablenotemark{b} & $1.54\pm0.04$\\
  R3 & $  44.9\pm0.5$ & $ 65.8\pm0.8$ & $337\pm6.4$ & $>2.57$\tablenotemark{b} & $2.00\pm0.04$\\
  R4 & $   8.0\pm0.4$ & $ 11.7\pm0.5$ & $241\pm6.4$ & 2.35 & $1.43\pm0.04$\\
  R5 & $  27.0\pm1.1$ & $ 39.5\pm1.7$ & $180\pm6.4$ & 1.08 & $1.07\pm0.04$\\
  R6 & $  39.3\pm1.3$ & $ 57.6\pm1.9$ & $166\pm6.4$ & $>0.81$\tablenotemark{b} & $0.98\pm0.04$\\
  R7 & $  27.9\pm1.3$ & $ 40.9\pm1.9$ & $107\pm6.4$ &   ---\tablenotemark{b} & $0.63\pm0.04$\\
  R8 & $  28.3\pm0.8$ & $ 41.5\pm1.1$ & $ 89\pm6.4$ & $>0.72$\tablenotemark{b} & $0.52\pm0.04$\\
  R9 & $  60.0\pm0.9$ & $ 87.9\pm1.3$ & $163\pm6.4$ & $>0.65$\tablenotemark{b} & $0.97\pm0.04$\\
 R10 & $  33.3\pm0.7$ & $ 48.8\pm1.0$ & $158\pm6.4$ & 0.77 & $0.94\pm0.04$\\
 R11 & $  51.7\pm0.3$ & $ 75.7\pm0.5$ & $174\pm6.4$ & 0.89 & $1.03\pm0.04$\\
 R12 & $  52.9\pm0.4$ & $ 77.5\pm0.6$ & $226\pm6.4$ & 0.97 & $1.34\pm0.04$\\
 R13 & $  41.8\pm0.8$ & $ 61.3\pm1.2$ & $212\pm6.4$ & 0.76 & $1.26\pm0.04$\\
 R14 & $  57.8\pm0.4$ & $ 84.6\pm0.6$ & $236\pm6.4$ & 1.21 & $1.40\pm0.04$\\
 R15 & $ 112.3\pm1.0$ & $164.5\pm1.5$ & $311\pm6.4$ & 2.17 & $1.84\pm0.04$\\
 R16 & $  90.3\pm1.8$ & $132.3\pm2.6$ & $381\pm6.4$ & 3.49 & $2.26\pm0.04$\\
 R17 & $  42.4\pm0.7$ & $ 62.1\pm1.0$ & $345\pm6.4$ & 2.81 & $2.05\pm0.04$\\
 R18 & $  47.8\pm0.8$ & $ 70.0\pm1.2$ & $172\pm6.4$ & 0.62 & $1.02\pm0.04$\\
 R19 & $  22.7\pm0.7$ & $ 33.2\pm1.0$ & $185\pm6.4$ & 1.15 & $1.10\pm0.04$\\
 R20 & $  50.7\pm0.6$ & $ 74.2\pm0.8$ & $207\pm6.4$ & 1.45 & $1.23\pm0.04$\\
 R21 & $  24.4\pm0.5$ & $ 35.7\pm0.8$ & $260\pm6.4$ & 1.33 & $1.54\pm0.04$\\
 R22 & $  13.9\pm0.4$ & $ 20.4\pm0.6$ & $231\pm6.4$ & 1.37 & $1.37\pm0.04$\\
 R23 & $  28.5\pm0.6$ & $ 41.7\pm0.9$ & $123\pm6.4$ & 0.27 & $0.73\pm0.04$\\
 R24 & $  38.0\pm1.4$ & $ 55.6\pm2.0$ & $105\pm6.4$ & 0.15 & $0.62\pm0.04$\\
\enddata
\tablecomments{Column (1): Region names. Column (2): Column
density. Column (3): Molecular gas mass. Column (4): Average
intensity of dust emission. Column (5): Average intensity of $\rm
H\alpha+[N II]$ emission. Column (6): Star formation rate surface
density.} \tablenotetext{a}{Image units in electrons per second
per pixels (eps~pixel$^{-1}$)} \tablenotetext{b}{These areas do
not have complete H$\alpha$ emission information, since these
regions are either located at the edge of or outside the
\emph{HST} image.}
\end{deluxetable}

In the spiral arm regions, the box sizes correspond to
$280\times350~$pc$^2$. The average values of $N_{\rm H_2}$ and
$M_{\rm H_2}$ per region are $39.8\times10^{21}$~cm$^{-2}$ and
$58.9\times10^6M_{\odot}$. The standard deviations of $N_{\rm
H_2}$ and $M_{\rm H_2}$ are large inside the spiral arms, with a
value of $24.0\times10^{21}$~cm$^{-2}$ and
$34.8\times10^6M_{\odot}$, respectively, indicating that the
molecular gas content within the spiral arms varies substantially
from region to region. On the other hand, $N_{\rm H_2}$ and
$M_{\rm H_2}$ for the central core C1 are estimated to be
$5.6\times10^{21}$~cm$^{-2}$ and $12.3\times10^6$~M$_{\odot}$
(note that the area of C1 is $\sim$1.5 times larger than those in
the spiral arms). This indicates that the column density and the
mass of the central region are much smaller than the average
values in the spiral arms. However, we note that the $R_{31}$ of
C1 is very different from those of the spiral arm regions (see
Sect.~\ref{sect-res-ratio}), and we have used different conversion
factors in estimating the mass and column density \citep{weis01}.

\citet{tac94} and \citet{stern94} showed that the $N_{\rm H_2}$
column density in the core region is about $4\times10^{22}$~
cm$^{-2}$ using the Galactic CO to $N_{\rm H_2}$ conversion factor.
Our conversion factor for the core region is about $\frac{1}{6}$ of
the Galactic conversion factor and we obtained a $N_{\rm H_2}$
column density of $5.6\times10^{21}$~cm$^{-2}$; in other words, the
different results are mainly caused by the different conversion
factors adopted. However, we note that a lower conversion factor is
typically found in galaxy centers \citep{wil95,mau96,weis01,esp09}.
In particular, the conversion factor in the center of NGC 1068 could
be six times lower than the Galactic value \citep{use04,gar10}.

\begin{figure}
\includegraphics[angle=-90,width=.9\textwidth]{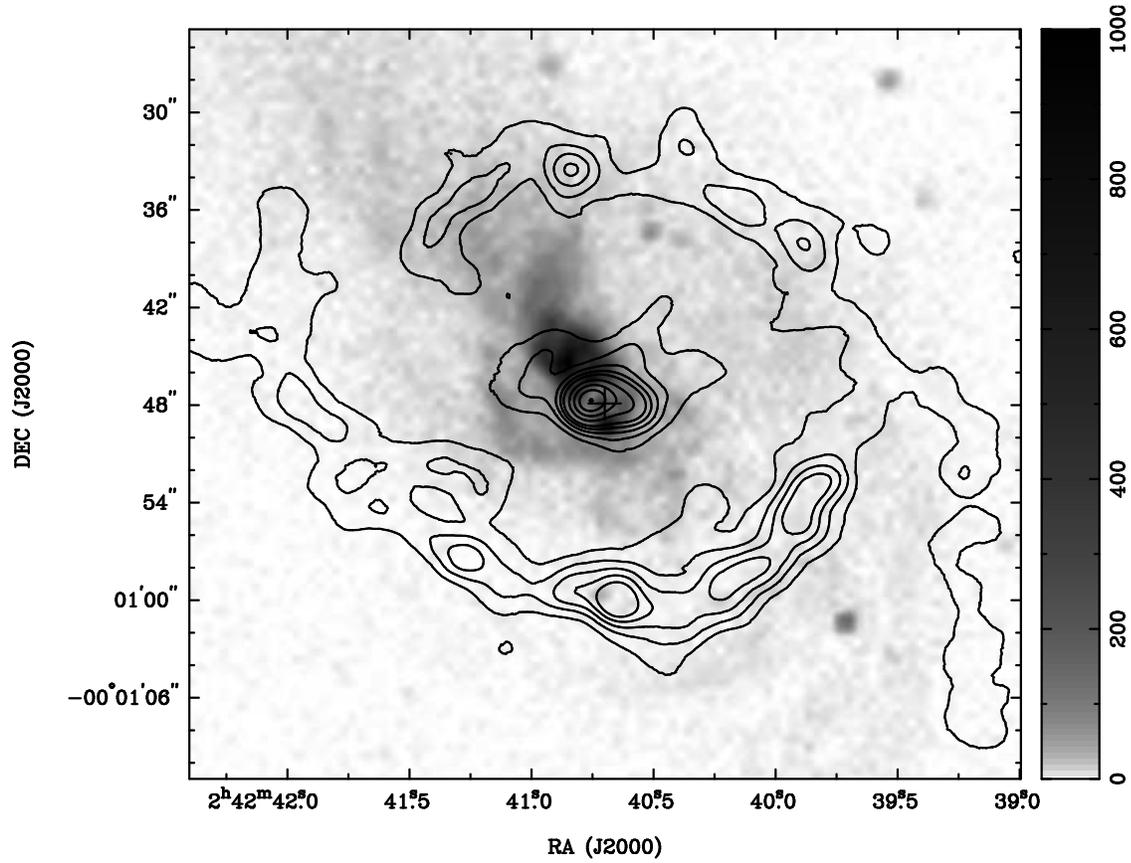}
\caption{Comparison between our CO(3--2) image and the X-ray image
    taken with the \emph{Chandra X-ray Observatory} \citep{you01}.
    The contours show CO(3--2) emission, with levels
    10, 20, 30, 40, 50, 70, 90, 110, and
    $130\times4.41$~Jy~beam$^{-1}$~km~s$^{-1}$.
    The gray scale shows the X-ray emission from 0 to 1000 counts pixel$^{-1}$($0.4-5.0$~keV).
    \label{fig5}}
\end{figure}

In Figure~\ref{fig5}, we compare our CO(3--2) image with the X-ray
image obtained with the \emph{Chandra X-ray Observatory}
\citep{you01}. The X-ray image displays clear emission from the
ionization cone emanating from the nucleus of NGC~1068; however, we
note that there is an obvious dimmed area at the center of the
image. This dimmed area matches well with the central core of our
CO(3--2) image, suggesting that the obscuring material of the X-rays
is closely related to the molecular gas.

\subsection{CO(3--2) Rotation Curve}
\label{sect-res-rot}

\begin{figure}
\includegraphics[angle=0,width=.9\textwidth]{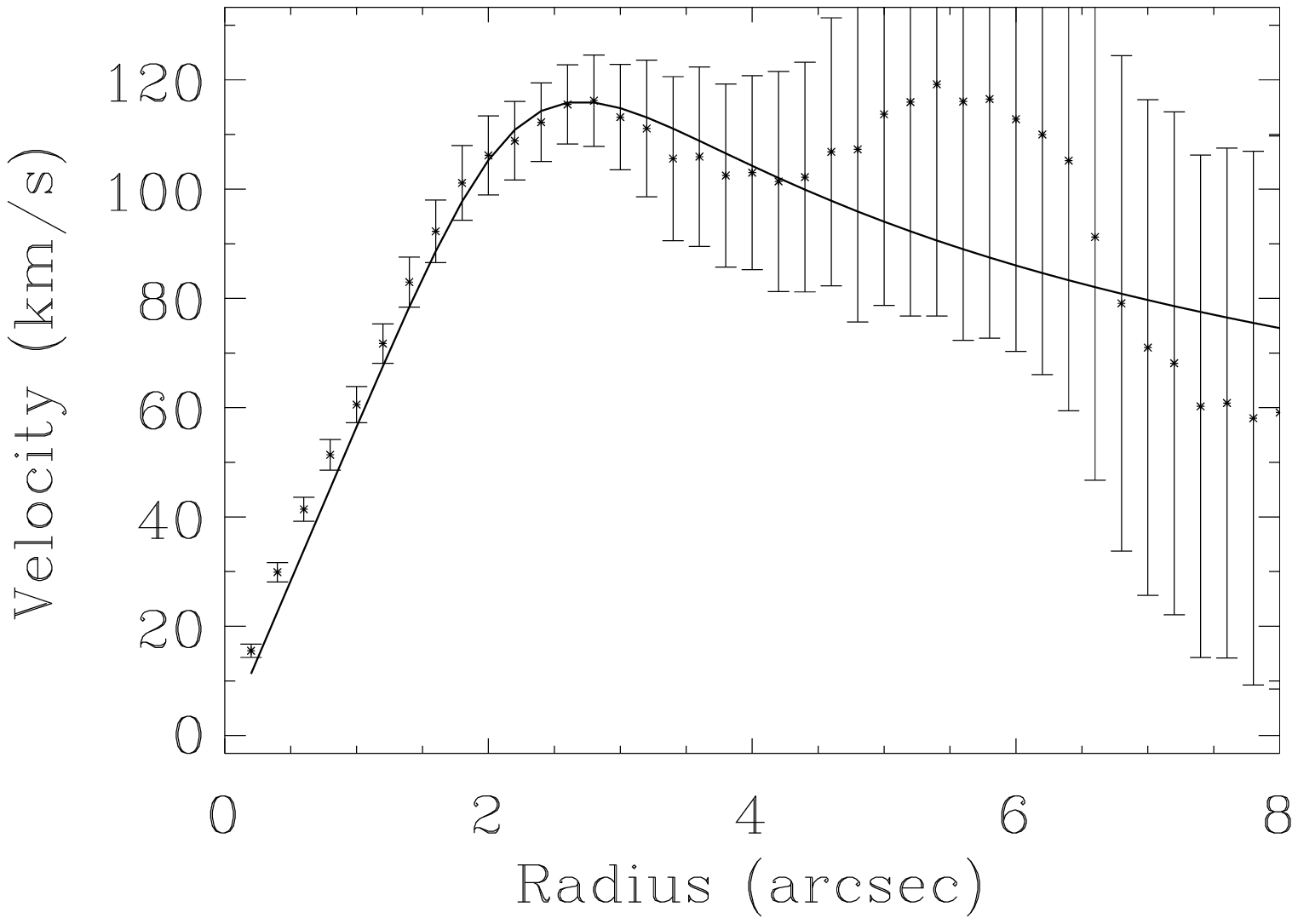}
\caption{Rotation curve of CO(3--2) in the central region of
NGC~1068. Crosses with error bars are the data points, and the
solid curve is the fitted Brandt rotation curve (see the main text
for details).
    \label{fig6}}
\end{figure}

The rotation curve in the nucleus of NGC~1068 is plotted in
Figure~\ref{fig6}. The data points and the error bars were derived
using the MIRIAD task VELFIT. We ignore points within $30\degr$ of
the minor axis (P.A.~$=60\degr$), and we assume the inclination
angle to be $45\degr$ following \citet{sch00}. Within the central
$2\arcsec$ ($\sim140$~pc), the rotational velocity increases with
radius (rigid rotation). The enclosed mass within the central
$2\arcsec$ can be estimated using $M(r<2\arcsec) \sim rv^{2}/G$, and
we find that the total mass within this area is
$3.7\times10^{8}$~M$_{\odot}$, which is consistent with
\citet{sch00}. The molecular gas mass in C1 is about
$12.3\times10^6$~M$_{\odot}$, so that the gas-to-dynamical mass
ratio is 3\%. This value is a factor of a few lower than those in
star forming galaxies ($\sim10\%$) \citep{sak99,kod05}.

For radii larger than $2\arcsec$, the rotation curve becomes
slowly decreasing. This rotation curve is well fitted with the
Brandt rotation curve \citep{bra60}:
\begin{equation}
V = \frac{V_{\rm max}\frac{R}{R_{\rm max}}}
    {\left(\frac{1}{3}+\frac{2}{3}
    \left(\frac{R}{R_{\rm max}}\right)^{n}\right)^{\frac{3}{2n}}}
\end{equation}
with $V_{\rm max}=116$~km~s$^{-1}$, $R_{\rm max}=2\farcs7$, and
$n=6$ (Figure~\ref{fig6}). The mass from the Brandt model is:
\begin{equation}
M_{\rm tot}
    = \left(\frac{3}{2}\right)^{\frac{3}{n}}
      \frac{V^2_{\rm max}R_{\rm max}}{G}.
\end{equation}
The total mass inside 190~pc ($\sim2.7\arcsec$) radius is
$7.2\times10^{8}$~M$_{\odot}$.

\subsection{Relation between CO Line Ratio and Star Formation}
\label{sect-res-sfr}

As shown in Sect.~\ref{sect-res-ratio}, $R_{31}$ varies along the
spiral arms. Since the CO(3--2) line is more closely related to
star formation \citep{kom07}, and $R_{31}$ increases with
increasing star formation efficiency \citep{mur07}, the variation
in $R_{31}$ along the spiral arms of NGC~1068 may also be related
to star formation. To study the cause of the variation, we compare
$R_{31}$ with star formation surface density along the spiral
arms.

We use both infrared and optical observations as star formation
tracers. The infrared data are obtained from the \emph{Spitzer Space
Telescope} IRAC $3.6~\micron$ and $8~\micron$
images\footnote{\url{http://ssc.spitzer.caltech.edu/archanaly/archive.html}}.
\citet{wu05} showed that dust emission at $8~\micron$ can be used as
a star formation indicator. We estimate the dust emission of in the
spiral arms of NGC~1068 using the observed \emph{Spitzer} IRAC
$3.6~\micron$ and $8~\micron$ fluxes:
\begin{equation}
f_{8\micron}(\mathrm{dust}) = f_{8\micron} - \eta_{8}
f_{3.6\micron},
\end{equation}
where $\eta_{8}=0.232$ \citep{lei99,hel04}. The derived values are
shown in Table~\ref{tab-sfr}.

H$\alpha$ emission is also considered as a tracer of star formation.
For NGC~1068, the wavelength of H$\alpha$ is shifted to 6587.8~\AA~
and those of [\ion{N}{2}] to 6573, 6603\AA. To obtain the H$\alpha$
emission, we retrieved the \emph{Hubble Space Telescope (HST)} F658N
and F791W images from the STScI
archive\footnote{\url{http://archive.stsci.edu}}. The F658N
narrow-band filter has a central wavelength of 6590.8~\AA~ with a
bandwidth of $\sim$30~\AA; therefore, the F658N image includes the
H$\alpha$ line as well as [\ion{N}{2}] lines, which are shifted to
6573 and 6603\AA for NGC~1068. Since this image also includes
continuum emission, we correct it using the nearby broad-band filter
F791W. The F791W filter has a central wavelength of 7881~\AA~ with a
bandwidth of $\sim$1231~\AA. The continuum subtraction can be
performed using the following equation:
\begin{equation}
f_{\rm line} = \frac{f_{\rm NB}\Delta\lambda_{\rm BB} - f_{\rm BB}\Delta\lambda_{\rm NB}}
    {\Delta\lambda_{\rm BB} - \Delta\lambda_{\rm NB}},
\end{equation}
where $f_{\rm line}$ is the continuum-subtracted line flux, $f_{\rm
NB}$ and $f_{\rm BB}$ are the observed total fluxes in the
narrow-band and broad-band filters, respectively, and
$\Delta\lambda_{\rm NB}$ and $\Delta\lambda_{\rm BB}$ are the
bandwidths of the narrow-band and broad-band filters, respectively.
The resulting continuum-subtracted F658N image can be considered as
the H$\alpha$ + [\ion{N}{2}] image. The derived values are shown in
Table~\ref{tab-sfr}. We cannot separate the H$\alpha$ from the
[\ion{N}{2}] lines and we cannot correct the H$\alpha$ +
[\ion{N}{2}] image for internal extinction with available data.
Therefore, we can not obtain any quantitive information from this
image; we only use the continuum-subtracted emission as a possible
indicator of relative star formation rates.

\begin{figure}
\includegraphics[angle=0,width=.9\textwidth]{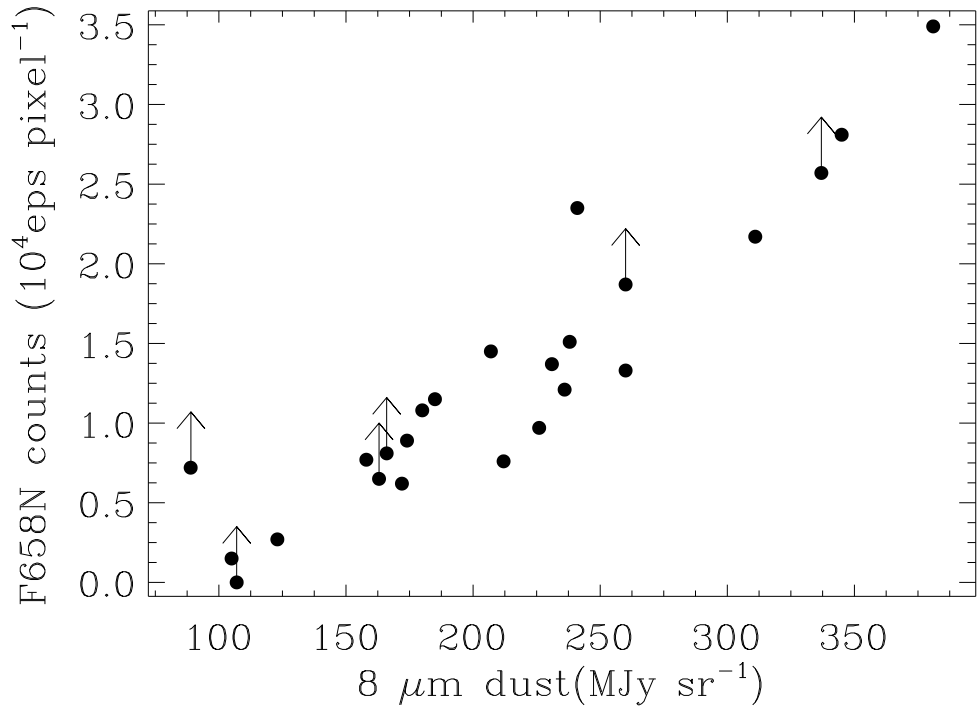}
\caption{Comparison of the \emph{Spitzer} $8~\micron$ dust
intensity with the \emph{HST} continuum-subtracted H$\alpha$ +
[\ion{N}{2}] line intensity. The \emph{HST} line intensity is not
flux-calibrated and is expressed in instrument units. The data
points with upper arrows do not have complete H$\alpha$ emission
information because they are located either at the edge of or
outside the \emph{HST} image. \label{fig7}}
\end{figure}

We compare the \emph{HST} H$\alpha$ + [\ion{N}{2}] line intensity
with the \emph{Spitzer} $8~\micron$ dust intensity.  As can be seen
in Figure~\ref{fig7}, the H$\alpha$ + [\ion{N}{2}] line intensity
and the $8~\micron$ dust intensity are linearly correlated,
indicating that the former image is effectively also a good star
formation tracer. We note that the CO distribution also matches very
well with the dust lanes along the spiral arms in the \emph{HST}
H$\alpha$ + [\ion{N}{2}] line image (Figure~\ref{fig3}). The star
forming regions are also located along the spiral arms but slightly
shifted toward the outside of the spiral arms, especially in the
southern spiral arm. This is similar to the results of previous
observations for the spiral arms of the nearby galaxy M51
\citep{vog88,aal99,kod09}.

\begin{figure}
\includegraphics[angle=0,width=.9\textwidth]{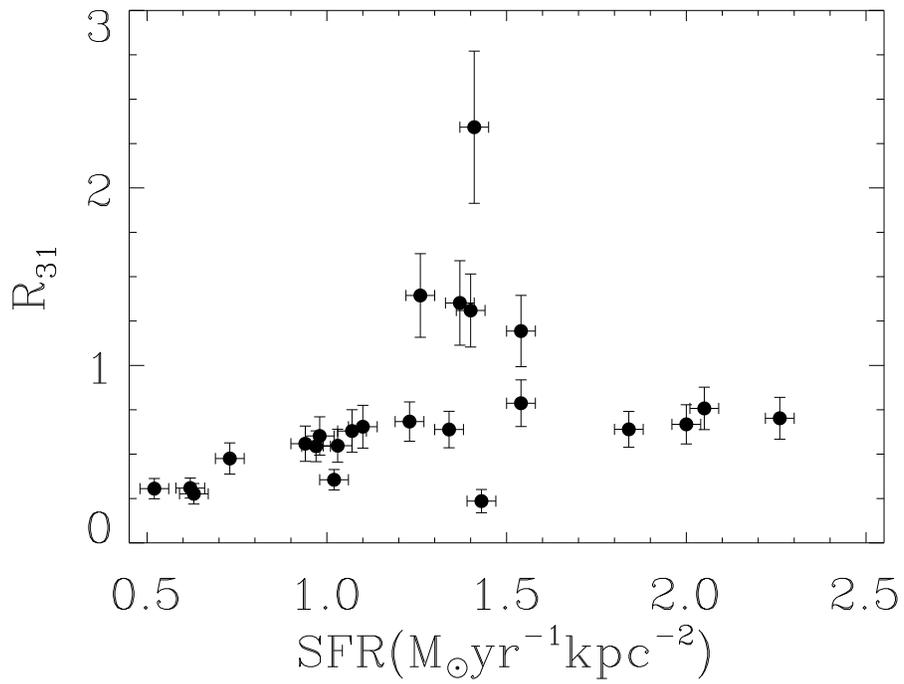}
\caption{Comparison between the CO(3--2)/CO(1--0)
    integrated intensity ratio, $R_{31}$, and the star
    formation rate surface density ($\Sigma_{\rm SFR}$). The error
    bars have included the $\sim~10\%$ uncertainty of our flux calibrations
    for the CO(3--2) and CO(1--0) emission.
    \label{fig8}}
\end{figure}

\begin{figure}
\includegraphics[angle=90,width=.9\textwidth]{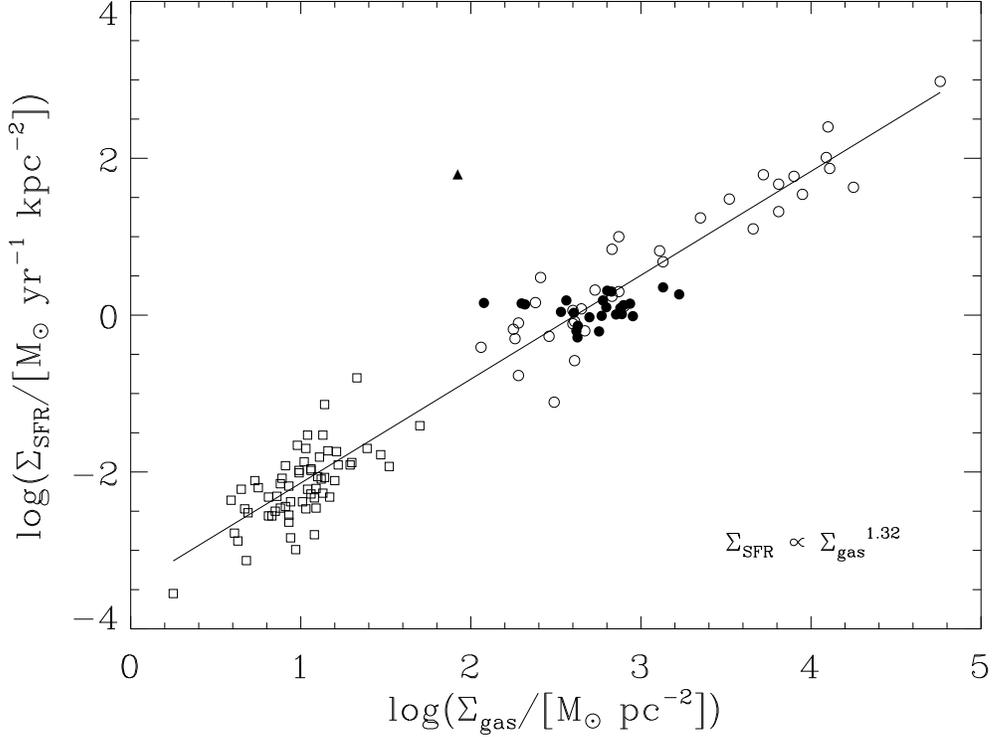}
\caption{Relation between the surface density of molecular gas
$\Sigma_{\rm gas}$ and $\Sigma_{\rm SFR}$ for normal galaxies
(open squares), starburst galaxies (open circles), and the spiral
arm regions of NGC~1068 (filled circles). The gas surface density
is derived from CO(1--0) emission. The normal and starburst galaxy
samples are obtained from \citet{ken98}. The solid triangle is the
nuclear region of NGC 1068 (C1). The solid line represents the
power-law fit to all data points. \label{fig9}}
\end{figure}

\begin{figure}
\includegraphics[angle=0,width=.9\textwidth]{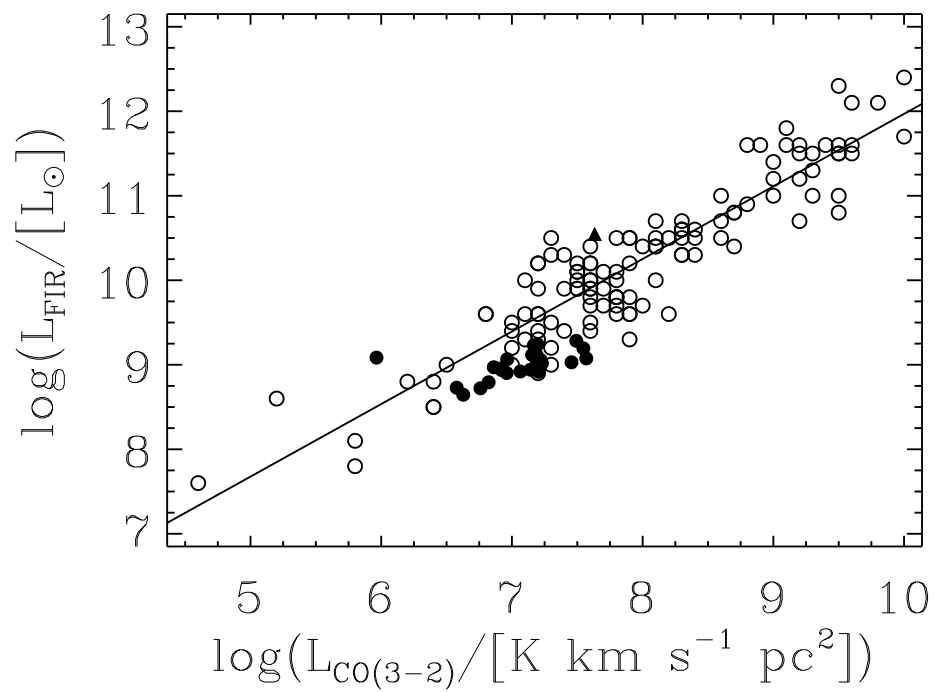}
\caption{Relation between integrated CO(3--2) and FIR
luminosities. The filled circles represent the data of the spiral
arm regions of NGC~1068, and the solid triangle is the nuclear
region of NGC 1068 (C1). The open circles are the data obtained
from \citet{mao10}. The solid line represents the linear fit to
all data points. The outlier point with the lowest CO(3--2)
emission is R4. \label{fig10}}
\end{figure}

We also compare molecular gas properties with the star formation
rate surface density ($\Sigma_{\rm SFR}$) derived from the
\emph{Spitzer} IRAC $8~\micron$ and $3.6~\micron$ data. The
$\Sigma_{\rm SFR}$ is derived from the $8~\micron$ dust
luminosity:
\begin{equation}
\frac{\Sigma_{\rm SFR}}{(M_\odot~{\rm yr^{-1}})} = \frac{\nu
L_{\nu}[8\micron(dust)]}{1.57\times10^9~{L_{\odot}}}
\end{equation}
\citep{wu05}, where $L_{\nu}$ is the $8~\micron$ dust luminosity.
The derived values are shown in Table~\ref{tab-sfr}.

Figure~\ref{fig8} shows the comparison between $R_{31}$ and
$\Sigma_{\rm SFR}$ of the spiral arm regions. The correlation
coefficient $r$ between the $R_{31}$ and $\Sigma_{\rm SFR}$ is 0.33
with a probability $p=89\%$ of mutual correlation. We note that most
of the $R_{31}$ of nearby star-forming galaxies are within 0.2--0.7
\citep{mau99}. For examples, \citet{mur07} found that the CO
emission of the starburst galaxy M83 has $R_{31} < 1$ and show a
good correlation between the $R_{31}$ and SFE of their data. If we
consider the data with $R_{31} < 1$ in our results, we find that the
correlation between the $R_{31}$ and $\Sigma_{\rm SFR}$ becomes
highly significant with $r=0.63$ and $p=99\%$. This suggests that in
normal star-forming regions the physical conditions of molecular gas
indicated by $R_{31}$ are well correlated with dust emission. The
data points where $R_{31}
> 1.0$ are obviously out of the correlation, suggesting that the
large $R_{31}$ might be caused by other reasons, such as a
different heating mechanism for the molecular gas or different
distributions of warm and cool molecular gas
\citep[e.g.,][]{ho87}.

Figure~\ref{fig9} shows the relation between the surface density of
molecular gas $\Sigma_{\rm gas}$ and $\Sigma_{\rm SFR}$. The star
formation rate surface densities of the spiral arm regions of NGC
1068 are much higher than those of normal galaxies and similar to
those of starburst galaxies; however, they all seem to follow the
same star formation law \citep{ken98}. This result strongly supports
the idea that the spiral arms in the inner $\sim2$ kpc region are
experiencing a starburst.  On the other hand, the molecular gas at
C1 is obviously offset from the Schmidt-Kennicutt law, suggesting
that C1 is mainly affected by AGN activities instead of star
formation.

In Figure~\ref{fig10}, we present the Schmidt-Kennicutt law for the
CO(3--2) emission. We used the FIR luminosity instead of the SFR in
Figure~\ref{fig9} so that it is easier to compare with the results
of \citet{mao10}. We first derive the star formation rate from the
observed $8\micron$ dust luminosity \citep{wu05} and then use the
SFR--$L_{FIR}$ relation, SFR$(M_\odot
yr^{-1})\sim1.7\times10^{-10}(L_{FIR}/L_\odot)$ \citep{ken98}, to
derive $L_{FIR}$. We note that this is effectively a
Schmidt-Kennicutt law for CO(3--2) emission since $L_{FIR}$ is
proportional to SFR, and the integrated CO(3--2) luminosity
represents the molecular mass in relatively warm and/or dense
regions. We find that most of our results follow the same relation
obtained by \citet{mao10} for nearby galaxies. The only outlier
point, which has the lowest CO(3--2) emission, is the interarm
region R4. When we combine our results with \citet{mao10}, we find
that the derived power-law index of the $L_{CO(3-2)}$--$L_{FIR}$
relation is $\sim0.9$. This value is smaller than the index of the
traditional Schmidt-Kennicutt law, which is around 1.0 to 2.0, but
is reasonable for the excitation conditions of warm and/or dense gas
\citep{kt07,na08}.

In Figure~\ref{fig9} and Figure~\ref{fig10}, the physical scale of
the NGC~1068 data is about $280\times350~$pc$^2$, which is much
smaller than those of \citet{ken98} and \citet{mao10}. We note that
there is no correlation between $\Sigma_{\rm gas}$ and $\Sigma_{\rm
SFR}$ in Figure~\ref{fig9} when only considering the NGC~1068 data.
On the other hand, the CO(3--2) emission from the same NGC~1068
regions show a very good correlation with the infrared as shown in
Figure~\ref{fig10}; the correlation coefficient $r$ is 0.789 with a
probability $p = 99.9\%$. The power-law index of the
$L_{CO(3-2)}$--$L_{FIR}$ relation for the NGC~1068 data alone
(excluding R4) is about 0.49. This value is significantly lower than
a typical power-law index of the Schmidt-Kennicutt law.

In Figure~\ref{fig10}, the CO(3-2) to FIR luminosity distribution is
generally consistent with previous studies, which have a power law
index of the Schmidt-Kennicutt law $\sim$1.0. On the other hand, the
Schmidt-Kennicutt law derived from the CO(1-0) data has a steeper
slope than that from the CO(3-2) data (Figure~\ref{fig9}).
Furtheremore, if we only consider our NGC 1068 data, we find a very
flat power law index, which cannot be explained by the model of
\citet{kt07} with a different critical density. One possibility is
that the gas is under sub-thermal conditions, which would produce a
flatter KS law as shown by \citet{na08}. This interpretation is also
consistent with the relatively large line ratios of
CO(3--2)/CO(1--0). We also note that most of our data are under the
average value of the KS law, indicating that the SFE of the inner
spiral regions of NGC 1068 is smaller than that in most of the
sources in Figure~\ref{fig10}.

There is also a radial variation of star formation activity in the
spiral arm regions. The average star formation rate surface density
of the inner arm regions (R1 -- R4, R9 -- R17, and R19 -- R22) is
1.45 $M_{\odot}$ yr$^{-1}$ kpc$^{-2}$, and the outer arm regions (R5
-- R8, R18, R23, and R24) is 0.80 $M_{\odot}$ yr$^{-1}$ kpc$^{-2}$.
In other words, the radial variation of the physical conditions of
the molecular gas mentioned in Sect.~\ref{sect-res-ratio} is
correlated with the radial variation of the galaxy's star formation.

\section{SUMMARY}
\label{sect-sum}

We have shown and compared the emission of different CO rotational
transitions of the prototypical Seyfert 2 galaxy NGC~1068 observed
with millimeter and submillimeter interferometers. The molecular
gas in the central part of this galaxy is distributed in a central
core and outer spiral arms. Both the CO(1--0) and CO(3--2) lines
show similar distribution along the spiral arms, and most of the
molecular gas mass is located in the spiral arms. However, the
nucleus is rather different; the strongest CO(3--2) peak lies in
the nucleus, but this is not true for CO(1--0). This is very
similar to another Seyfert 2 galaxy M51, suggesting that the AGN
is playing an important role in the different behaviors of these
two CO transition lines.

In the spiral arms, the CO(3--2)/CO(1--0) integrated intensity
ratio is well correlated with the star formation rate surface
density, indicating that the physical conditions of molecular gas
are related to star formation. Both the CO(3--2)/(1--0) ratio and
the star formation rate decrease with radius from the nucleus.

\acknowledgements

The authors thank an anonymous referee for important suggestions.
MT and CYH acknowledge support from the National Science Council
(NSC) of Taiwan through grant NSC 100-2119-M-008-011-MY3 and NSC
99-2112-M-008-014-MY3. SM acknowledges support from the NSC of
Taiwan through grant NSC 97-2112-M-001--021-MY3. DE was supported by
a Marie Curie International Fellowship within the 6$\rm^{th}$
European Community Framework Programme (MOIF-CT-2006-40298).


\end{document}